# Amorphous Mo-Ta oxide nanotubes for long-term stable Mo oxide based supercapacitors.

Bowen Jin[†, ⊥, ‡], Seyedsina Hejazi[⊥, ‡], Florian Pyczak[§], Michael Oehring[§], Shiva Mohajernia[⊥], Stepan Kment[#], Ondrej Tomanec[#], Radek Zboril[#], Nhat Truong Nguyen[⊥], Min Yang[†,*], Patrik Schmuki[⊥, #, ‖,*].

†MIIT Key Laboratory of Critical Materials Technology for New Energy Conversion and Storage, School of Chemistry and Chemical Engineering, Harbin Institute of Technology, Harbin, 150001, PR China.

⊥Department of Materials Science, Institute for Surface Science and Corrosion WW4-LKO, University of Erlangen-Nuremberg, Martensstraße 7, D-91058 Erlangen, Germany.

§Helmholtz-Zentrum Geesthacht, Zentrum für Material- und Küstenforschung GmbH, Max-Planck-Straße 1, 21502 Geesthacht, Germany.

#Regional Centre of Advanced Technologies and Materials, Palacky University Olomouc, 17. Listopadu 50A, 772 07 Olomouc, Czech Republic.

‖Department of Chemistry, King Abdulaziz University, Jeddah, Saudi Arabia.

**ABSTRACT:** With a large-scale usage of portable electric appliances, a high demand for increasingly high density energy storage devices has emerged. $MoO_3$ has, in principle, a large potential as negative electrode material in supercapacitive devices, due to high charge densities that can be obtained from its reversible redox reactions. Nevertheless, the extremely poor electrochemical stability of $MoO_3$ in aqueous electrolytes prevents a practical use in high capacitance devices. In this work, we describe how to overcome this severe stability issue by forming amorphous molybdenum oxide/tantalum oxide nanotubes by anodic oxidation of a Mo-Ta alloy. The presence of a critical amount of Ta-oxide (> 20 at-%) prevents the electrochemical decay of the $MoO_3$ phase and thus yields an extremely high stability. Due to the protection provided by tantalum oxide, no capacitance losses are measureable after 10000 charging/discharging cycles.

## 1. Introduction

The rapid growth of population and urgent energy requirements are promoting the development of renewable energy resources.[1] One crucial element in energy management systems and devices are supercapacitors that provide efficient energy storage and a high power density with a high charge/discharge rate.[2,3] With shrinking dimensions such as in portable electronics, or in electric vehicles, supercapacitors face new challenges and requirements. In addition to a high volumetric charge density[4–6] a high degree of stability during charging/discharging cycles is needed. Mainly 2D materials have been prepared and assembled to high volumetric energy density supercapacitors.[3,4,7,8] An alternative to 2D materials are aligned 1D nanostructures that may also provide a high surface area, but additionally have the advantage of a directional electron transport.[9] Moreover, many 1D nanostructured materials have ion transport geometries that ensure fast ion diffusion (and intercalation) processes.[10–12] Particularly interesting are nanostructured materials that not only rely on double layer charging, but additionally can contribute during charging/discharging cycles a significant pseudo-capacitance by redox reactions (i.e., materials with easily switchable oxidation states).

Typical examples are nanowires, rods and nanotubes of transition metal oxides, such as $WO_3$, $NiO$, $Fe_2O_3$, and $MoO_3$.[13–16] Among them, $MoO_3$ is energetically one of the most desirable negative electrode materials for its high theoretical capacitance.[17,18] However, the key obstacle to the use of $MoO_3$ in supercapacitors is the poor electrochemical stability in aqueous electrolytes especially acid solutions that prevents long-term cycling – as each red-ox cycle, and associated dissolution or structural collapse, degrades the material rapidly. Furthermore, the relatively poor conductivity of the pure oxide additionally limits the performance.[19,20] Therefore, various works reported on the fabrication of binary molybdenum oxide based compounds that aimed at combining a high capacitance with long-term cycling electrode material.[21,22] For example, Zhou et al have demonstrated the incorporation of tungsten oxide in $MoO_3$ will significantly improve its cycle ability and increase capacitive performance.[23] Mai et al. have demonstrated $MnMoO_4/CoMoO_4$ heterostructured nanowires with a reversibility with cycling efficiency of 98% after 1000 cycles.[24] Baskar etc. prepared α-$FeMoO_4$ as supercapacitor negative electrode material, the capacitance of the electrode remain 99% of the initial capacitance after 1000 cycles.[25] Yin et al. synthesized $NiMoO_4$ nanotubes by a hydrothermal treatment with $MoO_3$ nanorods. The structure showed a retention of 71% of the initial capacitance after 1000 cycles.[26] Senthilkumar et al. investigated $NiMoO_4$, $CoMoO_4$ and $MnMoO_4$ for supercapacitor materials application, the $NiMoO_4$ provided the largest reported capacitance of 1517 F/g but the material stability was only investigated for 200 cycles.[27] In other words, the key challenge in exploiting $MoO_3$ in supercapacitors remains the production of stable $MoO_3$ compounds that show a high capacitance and retain it over extended cycling.

In this work we describe an approach to intrinsically protect $MoO_3$ from dissolution by integrating Ta-oxide into a growing 1D nanotube structure. Tantalum oxide is well-known for its high electrochemical corrosion resistance in virtually any chemical environment and over a wide potential range.[28–30] To exploit this feature, we grow Mo-Ta oxide nanotubes with a sufficiently high Ta oxide content to provide maximum chemical/electrochemical stability, while maintaining a maximum capacitive switching performance provided by the Mo-oxide. Such a protection concept is borrowed from corrosion science where protection of an unstable species by a resistant species has been exploited for decades. The most prominent example are stainless steels (Fe-Cr alloys), where a critical amount of Cr drastically improves the electrochemical corrosion stability of the mixed oxide.[31] In general, protection of an unstable oxide by a stable oxide has been successfully described in the context of percolation theory.[32] Percolation models for 3D oxides that consist of a soluble and insoluble phase predict a protection threshold at 10-30 at-% of the corrosion resistant phase, to protect the entire oxide.[33,34]

Accordingly, here we use Mo-Ta alloys with 10-40 at-% Ta, to grow anodic self-ordered MoTaOx nanotube arrays to explore this threshold behavior. The anodic oxide nanotube layers form regular arrays that are directly back-contacted on the metallic alloy, i.e., they represent an electrode that is fabricated without additives or binder. This can provide be beneficial to the volumetric density and advantageous ion diffusion characteristics. These arrays then are explored as supercapacitor electrodes, not only for their performance but in particular for their cycling stability. Indeed, we find that a nanotube structure optimized for performance and stability is obtained for 20 at-% Ta alloy substrate. A thin nanotube layer delivers a volumetric capacitance up to 1484 F/cm³ (424 F/g) at 10 mV/s and 563 F/cm³ (160 F/g) at 200 mV/s. While these high capacitance values are related to the thinness of the layers most important is that a drastic improvement of the cycling stability is observed from <50 cycles for pure MoO₃ tubes to a retention of the capacitance at 100 % after 10000 cycles for the 20 at-% Ta alloy substrate.

## 2. Results and Discussion

In a first step, we investigated the feasibility to grow self-ordered oxide nanotubes on Mo-Ta alloys containing 10, 20, 30 and 40 at-% Ta. After various anodization parameters were optimized, we established that ordered nanotubular oxide layers can be established on all Mo-Ta alloys in an electrolyte consisting of 90 vol% glycerol and 10 vol% $H_2O$ containing 0.4 M $NH_4F$ and 0.1 M $(NH_4)_2SO_4$. Figures S1-S3 illustrate how various anodization parameters influence the nanotube morphology on the various Mo-Ta alloys. From this we deduced that using above electrolyte at 50 V and 25 °C yields overall comparable structures on the Mo as well as on the different Mo-Ta alloys. Figure 1a gives an example of the morphology of nanotubes on the Mo-20Ta alloy. A well-developed uniform nanotube layer with tubes of an individual length of 1.2 µm and a diameter of 130 nm was obtained. The tube length could be varied from 1-3 µm (Figure S2). Lower anodization temperatures formed thinner and nanoporous structures, and higher temperatures caused structural collapse. For water contents in the electrolyte lower than 5 vol% or higher than 20 vol%, either irregular nanoporous structure or thinner nanotube layers were formed. As mentioned, under optimized electrolyte conditions, nanotube layers could be formed on alloys with 10, 20, 30 and 40 at-% Ta. However, as different Ta-contents lead to different nanotube growth rates, a comparable morphology had to be established by adjusting the anodization times (shown in Figure S3).

As-formed nanotube layers for every composition as well as for pure Mo-oxide are amorphous (see XRD Figure 1c and Figure S4). After annealing in air at different temperatures (Figure 1b and Figure 1c), we find for all alloyed nanotubes that the morphology and amorphous nature is maintained up to 300 °C, while at higher temperatures (400 °C) an orthorhombic $MoO_x$ phase can be identified. It is noteworthy that at 300 °C pure $MoO_3$ nanotubes were converted to α-$MoO_3$, (JCPDS Card No. 05-0508). In other words tantalum induces an increase of the crystallization temperature, an observation which is in line with literature.[35] It should also be noted that the intensity of the XRD peaks originating from the alloy substrate decrease for the crystallized Mo-Ta oxides (treated at 400 °C or above). This suggests that the alloy substrate under these conditions is additionally thermally oxidized.[36] Indeed, Figure S5 shows a compact $MoTaO_x$ layer of a few hundred nanometers thickness underneath the $MoTaO_x$ nanotubes after annealing at 400 °C while lower temperatures only show a mild formation of a thermal oxide underlayer. In other words, if annealed at 400 °C we loose the advantage of a nanotube morphology that is directly connected to the metallic substrate.

To further characterize the ideal structure of the $Mo20TaO_x$ nanotubes after annealing at 300 °C, we carried out HRTEM, SAED and EDS mapping. Figure 1d shows HRTEM and SAED data for the tubes on the Mo20Ta alloy. The HRTEM image shows that no lattice fringes are apparent and only a diffuse halo ring is observed in SAED, confirming the amorphous nature of the nanotubes after annealing at 300 °C. The TEM/EDS-mapping confirms an entirely homogeneous distribution of Mo and Ta within the tube walls (as shown in Figure 1e) and the ratio of Mo and Ta from TEM-EDS is close to 4:1 which is the same as in the Mo-Ta substrate (shown in Figure S6).

This is further confirmed by X-ray photoelectron spectroscopy (XPS). Figure 2a shows the survey spectra of the amorphous tubes on the different alloys. From deconvolution of the high-resolution peaks, as shown in the Figure S7a and Figure 2c&d, the composition of the oxide was evaluated. The results of the Mo/Ta ratios are given in Figure 2b. These data further confirm that the amount of Mo and Ta in the nanotubes corresponds well to the substrate for every alloy. The XPS high resolution spectra for Ta in Figure 2c show spectra with a binding energy of Ta $4f_{7/4}$ and Ta $4f_{5/4}$ at 28.3 eV and 26.4 eV that corresponds to exclusively $Ta^{5+}$ in the oxide tube.[37] Figure 2d gives the XPS high resolution spectra for Mo in $MoO_3$ and $MoTaO_x$. The two spin orbit split peaks at 236.2 and 233.05 eV match well with data for pure $MoO_3$,[26,27] indicating molybdenum to be present as $Mo^{6+}$ in the $MoTaO_x$. It should be noted that a small peak located at 230.6 eV is apparent for $MoTaO_x$ nanotubes, which can be assigned to the $Ta3d_{5/2}$ orbit. The O1s high resolution spectrum in Figure S7a can be separated into two peaks at 531.02 and 530.7 eV, corresponding to individual O-$Mo^{6+}$ and O-$Ta^{5+}$ species, respectively.[26,28] This further confirms the presence of individual oxides, i.e. that the $MoTaO_x$ amorphous nanotubes consist of a mixture of $Ta^{5+}$-oxide and $Mo^{6+}$-oxide units rather than any type of a binary oxide. Even after crystallization, the nanotubes still show the characteristics of a mixture (Figure S7b&c). This in line with XRD (Figure 1c) where also the crystallized $Mo20TaO_x$ nanotubes still show the features of the individual oxides; no binary oxide compound can be identified.

The electrochemical behavior of all nanotube-electrodes was first assessed by CV measurements to establish a suitable potential window for the $MoTaO_x$ electrodes, as illustrated in Figure S8. This window was selected to avoid electrolyte decomposition while maintaining a high capacity during long-term cycling. It should be noted that an amorphous $MoO_3$ nanotube film dissolved in 1 M $H_2SO_4$ aqueous solution in 20 minutes, even without applied voltage (as shown in Figure S9), while crystallized $MoO_3$ nanotubes can be tested in this potential window but only for a few cycles. Figure 3a shows CV plots of $MoTaO_x$ nanotubes with different Ta contents, indicating very similar characteristics for each Ta content. From the CV curves, the volumetric capacitance was obtained by integrating the area within the CV (Figure 3b). Clearly, the highest capacitance values are obtained for the $Mo20TaO_x$. It is noteworthy that the capacitance of $Ta_2O_5$ is less than 1 F/cm³ at all scan rate due to the poor conductivity and high electrochemical inertness of pure $Ta_2O_5$. For pure $MoO_3$, the first CV shows, as expected, a very high capacitive charge, but this can only be observed for one cycle (the CV plots of crystallized $MoO_3$ in initial three cycles are shown in Figure S10). For all Ta/Mo-oxide electrodes, to reach a maximum capacitance, an initial cycling was required. As outlined in the supporting information (Figure S11), in this first phase, conductive hydrogen molybdenum bronze ($H_xMo20TaO_x$) is formed that leads to a steady increase in the capacitance (up to ≈1000 cycles).

After these "initiating cycles" the Mo20TaO$_x$ electrode exhibits the highest specific capacity at all scan rates with ~1484 F/cm$^3$ (424 F/g) at 10 mV/s, which is almost twenty times larger than found in the initial cycles for the crystallized MoO$_3$ electrode (67 F/cm$^3$). The favorable rate performance of Mo20TaO$_x$ still remains even for thinner or thicker layers (as shown in Figure S12). For Mo20TaO$_x$ with a tube length of 3.3 µm (the mass of thicker film is ~1.2 mg/cm$^2$), the capacitance achieves 1390 F/cm$^3$ at 10 mV/s, which is comparable to that of other non-noble metal electrodes of a similar thickness (a comparison to literature is given in Table S2). Galvanostatic charge-discharge (GCD) curves of Mo20TaO$_x$ electrodes at different current densities are shown in Figure 3c. The GCD plots of Mo20TaO$_x$ are asymmetric with different current densities. We suggest that the asymmetry is provided by a multi-electron reaction process. The reaction mechanisms in literature [38-40] suggests that H$^+$ are intercalated successively and the intermediate products in the oxidation and the reduction process are different (see supporting information), resulting in a slight deviation in the symmetry of the GCD plots. The discharge time of Mo20TaO$_x$ is 227.2 s at 5 A/cm$^3$ and 5.9 s at 100 A/cm$^3$, which is ten times longer than for crystallized MoO$_3$ and million times longer than Ta$_2$O$_5$. By calculating the volumetric capacitance from GCD curves, the Mo20TaO$_x$ electrode reaches 1893 F/cm$^3$ (541 F/g) at 5 A/cm$^3$ (~1.5 A/g), and maintains 893 F/cm$^3$ (280 F/g) at 100 A/cm$^3$ (~30 A/g).

Figure S13a shows that the characteristics of the CV plots are strongly dependent on whether or not the Mo20TaO$_x$ is crystalline. For the crystallized Mo20TaO$_x$, three pairs of peaks are observed, indicating that 1.6 electron equivalent reversible Faradic reactions take place on the MoTaO$_x$ electrode. In contrast, no redox peaks are observed in the CV curves of the amorphous Mo20TaO$_x$ in the same potential window. This absence of distinct intercalation peaks has also been reported for other amorphous materials.[18,41] Figure S13b shows that the crystallized Mo20TaO$_x$ only reaches a volumetric capacitance ~1109 F/cm$^3$ (mass capacitance of 254 F/g) at the scan rate of 10 mV/s and 312 F/cm$^3$ (71 F/g) at 200 mV/s. In contrast, amorphous Mo20TaO$_x$ delivers ~1484 F/cm$^3$ (424 F/g) at 10 mV/s and 563 F/cm$^3$ (160 F/g) at 200 mV/s. Clearly, the amorphous Mo20TaO$_x$ exhibits a higher capacitance than the crystallized Mo20TaO$_x$. This can be ascribed to the structurally disordered nature of amorphous phases that can better accommodate expansion or contraction during ion intercalation and diffusion process.[42,43] Another reason for the higher capacitance of the amorphous nanotubes is that, as mentioned, crystallization of the Mo20TaO$_x$ nanotubes at 400 °C is accompanied by formation of a compact oxide layer under the nanotubes. The compact layer not only increases the thickness and mass of the active layer of the electrode, but also leads to a loss of some advantages of the nanotubes, such as a high specific surface area or a fast ion diffusion. The capacitance performance of crystallized Mo20TaO$_x$ therefore is significantly decreased by the presence of the compact layer, especially for higher scan rates.

However, the key point of our supercapacitor concept is the cycling stability. For this, we evaluated the long-term cyclic performance by continuous CV measurements at scan rate of 100 mV/s. Figure 4a illustrates the capacitance retention behavior during cycling for the different Mo/Ta-oxide electrodes and pure MoO$_3$. Although crystallized MoO$_3$ has a good capacitance performance in the initial cycles, the capacitance quickly decreases in the following four cycles; only 21 F/cm$^3$ remain after 1000 cycles. Optical images show that the active film of MoO$_3$ partly peeled off after 1000 cycles. SEM images indicate that the nanotube morphology was lost and only pits were present in the substrate after the continuous cycling (shown in Figure S14a&b). The corresponding XRD in Figure S14c show that the main signals of the XRD pattern are Mo substrate peaks and only weak signal of α-MoO$_3$ can be detected after cycling. In contrast, the addition of Ta can drastically improve the cyclic stability of the electrode. Figure 4a shows that the capacitance of Mo10TaO$_x$ (Mo:Ta 9:1) remains at 93.5 F/cm$^3$ (~20 % of initial capacitance) after 10000 cycles. All higher amounts of Ta even more drastically improve the cyclic stability. This is highly beneficial for tubes containing 20 and 30 at-% Ta. For tubes containing 40 at-% of Ta, the tubes are stable but the high Ta content leads to a drastic drop in capacitance. In other words, a maximum synergy in terms of capacity and stability is obtained for Mo20TaO$_x$ that exhibits an outstanding stability, with almost no measurable capacitance losses even after 10000 cycles. Moreover, the coulombic efficiency is almost 100% during the cycling. Nyquist and Bode plots do not show any obvious changes during the cycling process (as shown in Figure S15), illustrating the high stability in the capacitive behavior. To our knowledge, the cycling stability is the best ever reported up to date for any MoO$_3$-based supercapacitor negative electrodes (please see comparison in Table S3).

The major issues for MoO$_3$ related to the stability of a supercapacitor negative electrode (in acid aqueous media) are the strong dissolution,[44-46] poor structural stability,[38] and irreversible phase transition[47] during the charging process. Ta oxide was selected for protection as it provides an outstanding corrosion resistance in virtually all environments. As a result, the addition of Ta oxide almost completely suppressed chemical and electrochemical-corrosion of the 2-component oxide leading to the very high stability in acid solution (see Figure S9) [this in analogy to Cr-oxides protecting Fe-oxide in the passive film of stainless steel].[48,49] Additionally relevant to intercalation devices amorphous materials have a higher flexibility to tolerate structural distortion and avoids the 1D structure collapse during a charge storage process. The excellent cycling performance can also attributed to the amorphous nature. In general, amorphous materials have a higher flexibility to tolerate structural distortion and avoids the 1D structure collapse during a charge storage process.[50] Accordingly post-mortem SEM shows no structural collapse of the tubular features after cycling (shown in Figure 4b), and the corresponding XRD show amorphous character as before cycling (shown in Figure S16). XPS results in Figure 4c illustrate that only small variations were observed in the Mo valence state after long-term cyclic performance, manifesting a good reversibility of Mo20TaO$_x$. Moreover, no changes did occur in the Ta peak after cycling, (shown in Figure 4d), confirming the outstanding stability of Ta oxide. Evidently a level of 20 at-% Ta oxide can protect the entire amorphous nanotube structure during extended cycling – this in line with thresholds predicted by 3D percolation models.[33,34]

In a wider context, our work provides a new strategy to fabricate highly stable electrodes from materials with high capacity but low stability. By forming an amorphous mixture with a stable secondary compound, in our case Ta oxide, an extremely high electrochemical stability of the primary functional material can be reached.

## 3. Conclusions

In summary, we demonstrate a strategy to establish an extremely long cycling life-time in $MoO_3$-based supercapacitor electrodes. Key is to establish an amorphous composite with an extremely stable oxide, which in our case is $Ta_2O_5$. As $Ta_2O_5$ is electrochemically and chemically inert, it is able to form, in a sufficient concentration, a scaffold that protects the entire oxide layer from degradation (similar to the percolative protection on stainless steels where sufficient Cr-oxide protects susceptible Fe-oxide). In our case, we grow amorphous oxide nanotubes on Mo-Ta alloys. At a critical concentration >20 % Ta-oxide in the MoTaOx chemical and electrochemical degradation of the 2-component oxide is completely suppressed. In spite of the addition of the inert oxide, the electrode with 20 % Ta oxide shows a remarkable efficiency surpassing pure $MoO_3$ after a few cycles. Once amorphous MoTaOx is fully activated by initial cycling (forming conductive Mo-bronze in the composite), a capacity of 1484 $F/cm^3$ (424 F/g) at 10 mV/s and 563 $F/cm^3$ (160 F/g) at 200 mV/s can be reached and most importantly: maintained during cycling. This extremely high cycling stability, is best illustrated for the Mo20TaOx electrode where no measurable capacitance loss even after 10000 cycles is observed.

## 4. Experimental Section

### Fabrication of MoTaO$_x$ nanotubes

MoTaO$_x$ nanotubes were grown from Mo-Ta (Mo:Ta = 9:1, 8:2, 7:3 and 6:4) alloys by anodization. Prior to anodization, Mo-Ta alloy samples were degreased by ultrasonication in acetone, ethanol and distilled water successively, followed by drying under a nitrogen stream. Ta-Mo alloys were anodized in glycerol with addition of 0.4 M $NH_4F$ and 5 vol% ~30 vol% $H_2O$ electrolyte at 50 V at different temperatures with thermostat. The Ta-Mo alloy served as the working electrode and platinum electrode was used as a counter electrode. After anodization the samples were immersed in ethanol to remove residual $F^-$ and then dried under a $N_2$ stream. The samples were annealed at 200 ~ 600 $^o$C in air for two hours to transform the amorphous nanotubes to a crystalline structure.

### Material Characterization

The morphology of the MoTaO$_x$ and $MoO_3$ nanotubes was characterized by scanning electron microscope (SEM Hitachi S4800). The crystal structure was examined by X-ray diffraction (XRD Cu Kα radiation, pert Philips MPD). The chemical composition was analyzed by X-ray photoelectron spectroscopy (XPS, PHI 5600, US).

### Electrochemical measurements

Cyclic voltammetry (CV), electrochemical impedance spectroscopy (EIS), and galvanostatic charge-discharge (GCD) measurements were carried out with a three electrode system, MoTaO$_x$ or $MoO_3$ were used as a working electrode, graphite rod was used as a counter electrode and AgCl/Ag (3M KCl) as a reference electrode. CV and EIS were performed with a Zahner electrochemical workstation. GCD was measured by Autolab (PGSTAT30) electrochemical workstation. All the measurements were performed in 1M $H_2SO_4$ at room temperature. Prior to the electrochemical measurements, the MoTaO$_x$ nanotube layers were electrochemically activated for one thousand CV cycles with a scan rate of 100 mV/s. The impedance measurement of the sample was measured in the frequency range from 100 kHz to 10 mHz with a potential perturbation of +/-10 mV amplitude on the open circuit potential.

Assembly of the ASC Device:

The ASC device was assembled by two electrodes, using Mo20TaO$_x$ as the negative electrode and activated carbon electrode was used as the positive electrode with a 1M $H_2SO_4$ electrolyte.

### Calculations

The area capacitance of each sample is calculated from the CV according to the following equation[51]:

$$C_s = \frac{1}{2} \times (\frac{1}{A(V_2 - V_1)} \frac{dV}{dt} \int_{V_1}^{V_2} I_{charge} dV + \frac{1}{A(V_2 - V_1)} \frac{dV}{dt} \int_{V_2}^{V_1} I_{discharge} dV)$$

where $C_s$ is area capacitance, A is the exposure area of each sample (c.a. 0.672 $cm^2$), I is current, dV/dt is scan rate, $V_1$ and $V_2$ are potential limit of CV measurement.

The volumetric capacitance is calculated by the use of following equation

$$C_v = \frac{C_s}{d}$$

where d is the thickness of the film.

The thickness of the film is determined from the nanotube length from the cross-sectional SEM images. The average thickness of MoTaO$_x$ is approximately 1.2 μm and the volume of active material is about 8.0×$10^{-5}$ $cm^3$.

The volumetric capacitance measured by GCD plots is calculated according to the following equation[51]:

$$C_v = \frac{I}{V} \frac{t_{discharge}}{(V_1 - V_2)}$$

where I is the set current, V is the film volume, $t_1$(s) and $t_2$(s) are charge and discharge time respectively and $V_1$(V) and $V_2$(V) are the GCD potential limited.

The volumetric capacitances and specific capacitances of the electrodes were also calculated based on GCD curves as follows:

$$C_m = \frac{I}{m_{MoTaOx}} \frac{t_{discharge}}{(V_1 - V_2)}$$

### Estimation of the active layer weight

The weight of the active layer was measured by following method:

The mass of the Mo20TaO$_x$ active layer was measured by a high resolution balance. After weighing, the samples were immersed in 1M KOH for 24 hours and then ultra-sonicated for a few seconds. After that substrate was cleaned by distilled water and measured again. To reduce the errors, the sample was measured 3 times. Overall, the average mass for amorphous Mo20TaO$_x$ nanotube layer is 0.28 mg and the crystallized Mo20TaOₓ nanotube layer is 0.55 mg.

## ASSOCIATED CONTENT

The Supporting Information is available free of charge on the ACS Publications website.

## AUTHOR INFORMATION

**Corresponding Author**

E-mails: schmuki@ww.uni-erlangen.de, yangmin@hit.edu.cn.

**Author Contributions**

‡These authors contributed equally.

## ACKNOWLEDGMENT

The authors would like to acknowledge ERC, DFG, and the DFG cluster of excellence "Engineering of Advanced Materials" for financial support. B.J. and M.Y. thank the National Natural Science Foundation of China (No.21773046), B.J. thanks the China Scholarship Council (CSC) for financial support. The authors thank also Dr. Ning Liu, Imgon Hwang, Anja Friedrich for SEM measurement, Helga Hildebrand for XPS analyses and Nikita Denisov for TEM measurement.

Figures

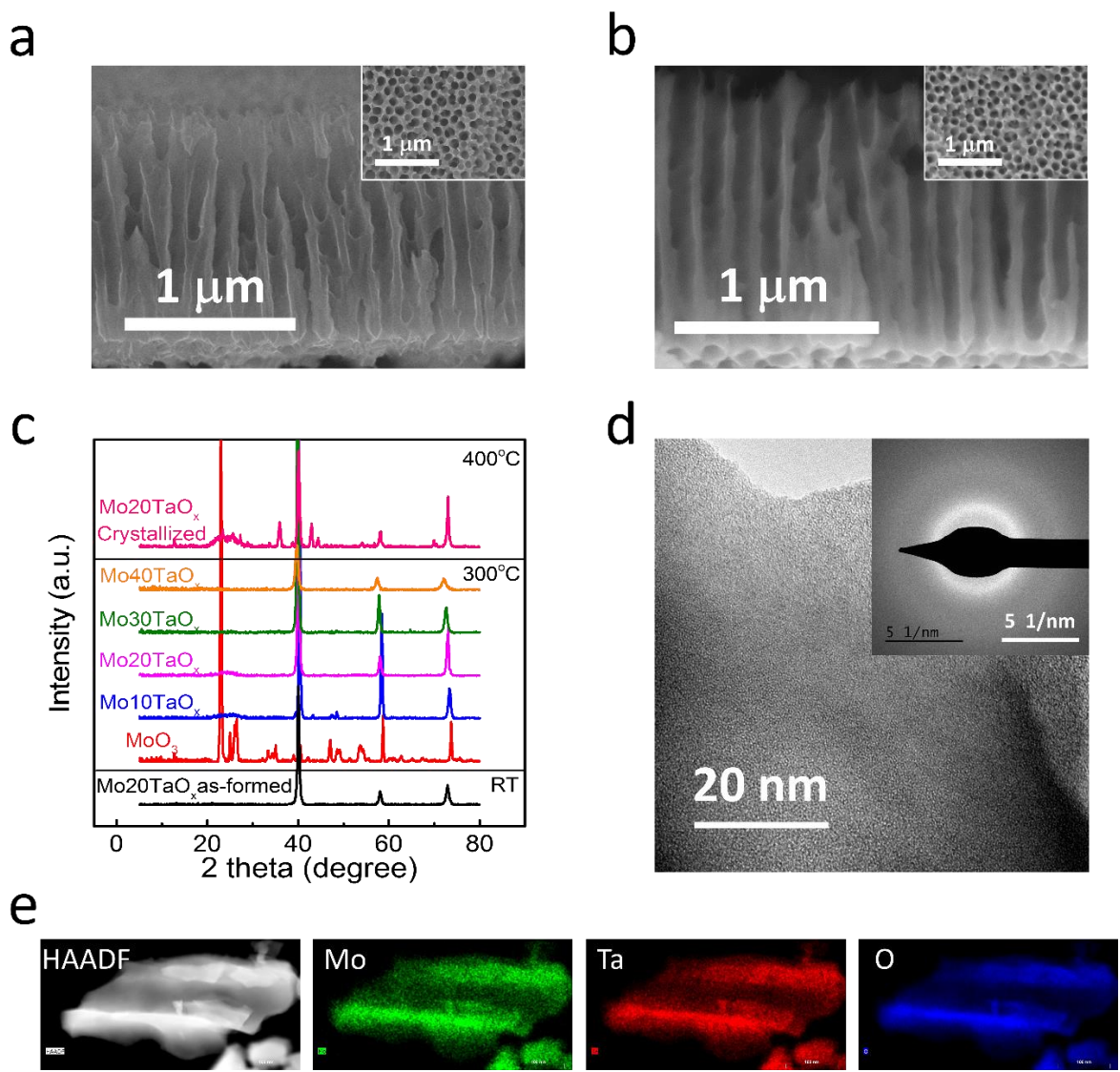

**Figure 1** (a) SEM image of as-formed Mo20TaO$_x$ nanotubes anodized at 50 V (inset: cross-section of nanotubes); (b) SEM image of Mo20TaO$_x$ nanotubes after annealing in air at 300 ℃ (inset: cross-section of nanotubes); (c) XRD diffractions of amorphous and crystallized MoTaO$_x$; (d) HRTEM image of amorphous Mo20TaO$_x$ nanotubes after annealing (inset: SAED of amorphous Mo20TaO$_x$ nanotubes); (e) TEM-EDS map of Mo, Ta, O for amorphous Mo20TaO$_x$ nanotubes.

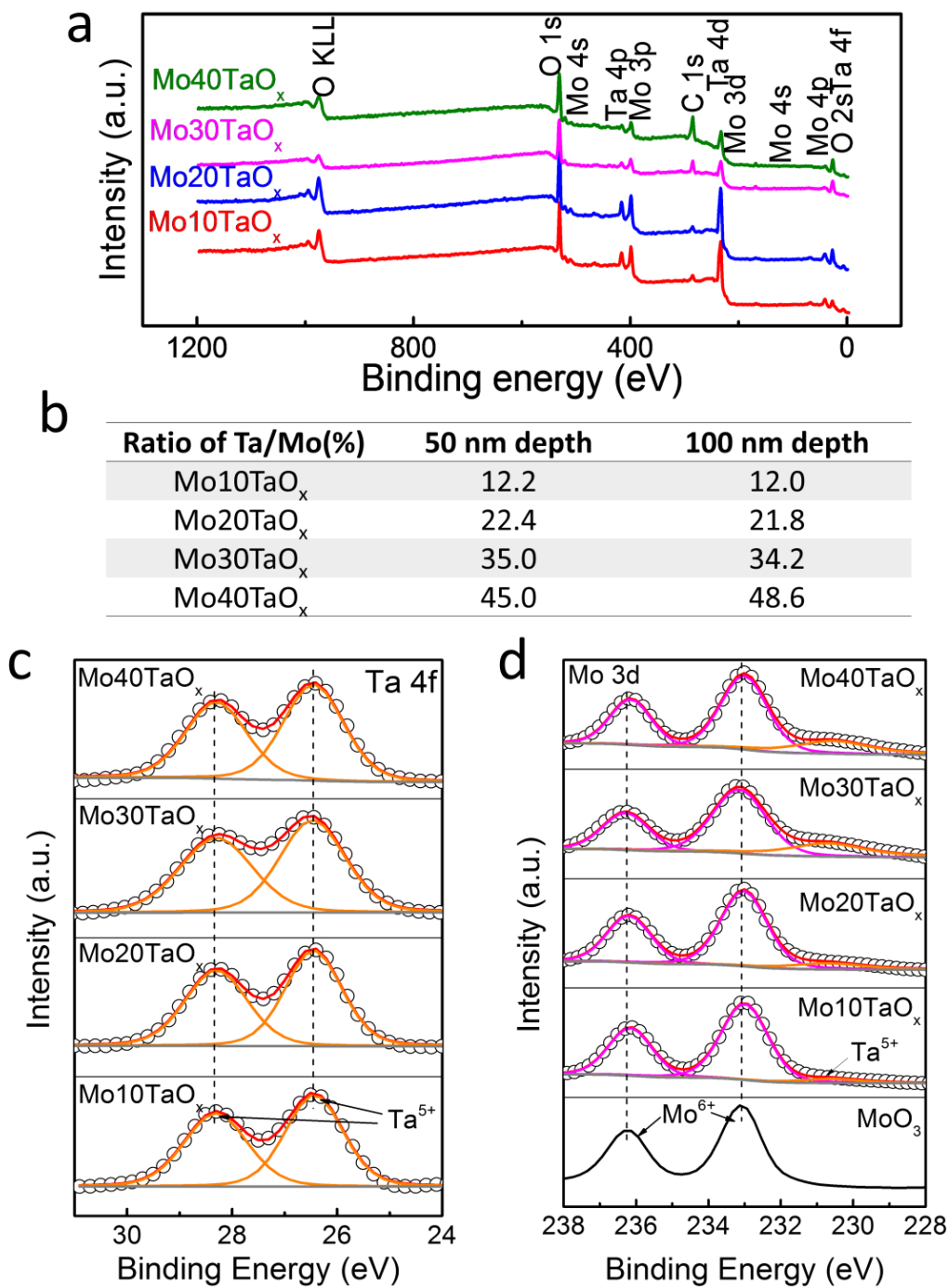

| Ratio of Ta/Mo(%) | 50 nm depth | 100 nm depth |
|---|---|---|
| Mo10TaO$_x$ | 12.2 | 12.0 |
| Mo20TaO$_x$ | 22.4 | 21.8 |
| Mo30TaO$_x$ | 35.0 | 34.2 |
| Mo40TaO$_x$ | 45.0 | 48.6 |

**Figure 2** Composition of MoTaO$_x$ nanotubes as determined by XPS: (a) XPS survey spectras of MoTa oxide with different Ta concertration; (b) atomic concentration of elements in 50nm and 100nm depth; (c) Ta 4f high resolution spectra; (d) Mo 3d high resolution spectra.

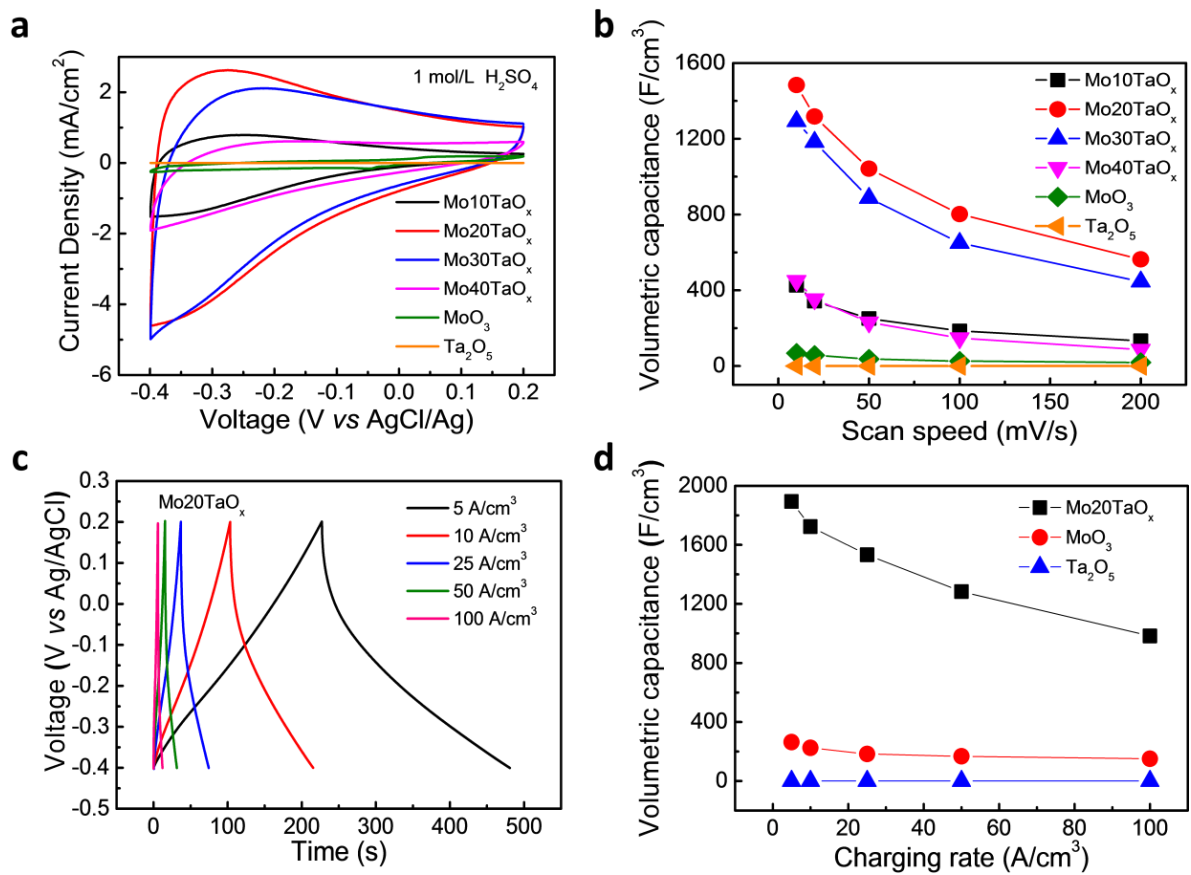

**Figure 3** Electrochemical characterization: (a) CV plots of MoTaO$_x$ and MoO$_3$ with scan rate of 10 mV/s; (b) volumetric capacitance of MoTaO$_x$ and MoO$_3$ at different scan rates determined from the CV plots; (c) galvanostatic charge-discharge plots of Mo20TaO$_x$ at different charging/discharging rates; (f) volumetric capacitance of Mo20TaO$_x$ and MoO$_3$ at different charging/discharging rates.

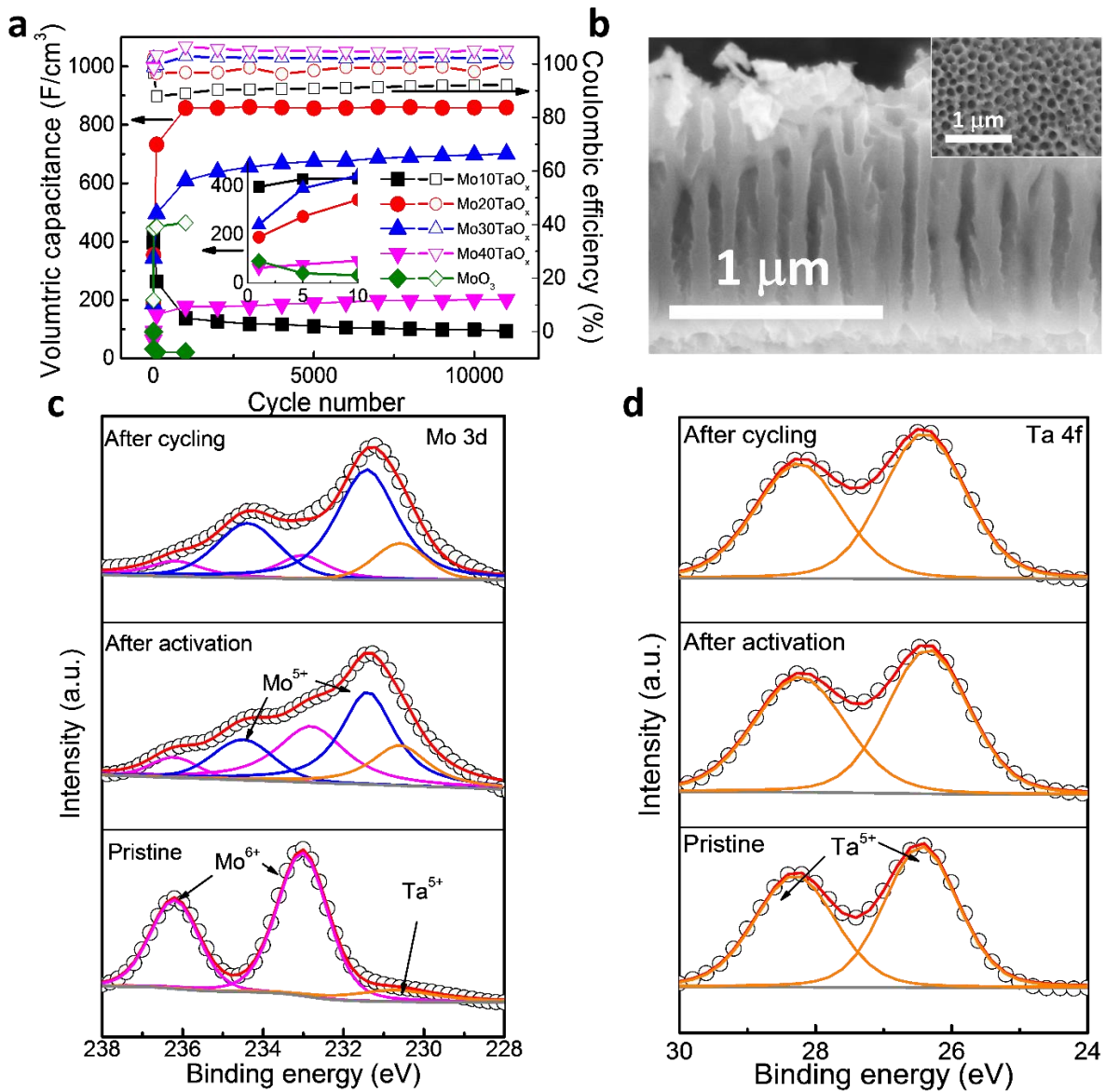

**Figure 4** Stability characterization: (a) Capacitance retention of MoTa oxide with different Ta concertation and MoO₃ during cycling; (b) SEM image of Mo20TaOₓ after cycling; (c) high resolution XPS spectra of Mo 3d spectra for as-formed Mo20TaOₓ, after activation and after 10000 cycles; (d) high resolution XPS spectra of Ta 4f for as-formed Mo20TaOₓ, after activation and after 10000 cycles.

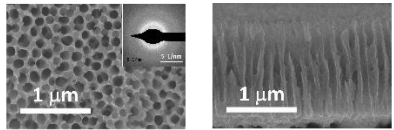

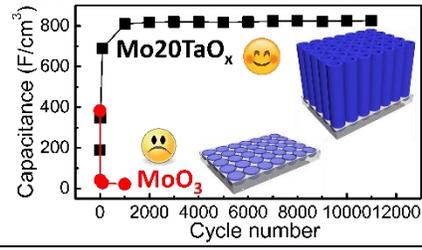